\documentclass[aps,pra,showkeys,showpacs]{revtex4}
\usepackage{amsmath}
\usepackage{bm}
\usepackage{epsfig}

\begin{document}

\title{Quantum random walks with history dependence}
\author{A. P. Flitney}
\affiliation{Centre for Biomedical Engineering (CBME)
and Department of Electrical and Electronic Engineering, \\
The University of Adelaide, SA 5005, Australia}
\email{aflitney@eleceng.adelaide.edu.au, dabbott@eleceng.adelaide.edu.au}
\author{N. F. Johnson}
\affiliation{Centre for Quantum Computation and Physics Department, \\
Clarendon Laboratory, Oxford University, Parks Road, Oxford, OX1 3PU, U.K.}
\email{n.johnson@physics.ox.ac.uk}
\author{D. Abbott}
\affiliation{Centre for Biomedical Engineering (CBME)
and Department of Electrical and Electronic Engineering, \\
The University of Adelaide, SA 5005, Australia}
\email{aflitney@eleceng.adelaide.edu.au, dabbott@eleceng.adelaide.edu.au}
\date{\today}

\begin{abstract}
We introduce a multi-coin discrete quantum random walk
where the amplitude for a coin flip depends upon previous tosses.
Although the corresponding classical random walk is unbiased,
a bias can be introduced into the quantum walk
by varying the history dependence.
By mixing the biased random walk with an unbiased one,
the direction of the bias can be reversed
leading to a new quantum version of Parrondo's paradox.
\end{abstract}

\pacs{03.67.-a, 05.40.Fb, 02.50.Le}
\keywords{quantum random walks, Parrondo's games,
quantum lattice gas automata, non-Markovian dynamics}

\maketitle

\section{Introduction}
Random walks have long been a powerful tool in mathematics,
have a number of applications in theoretical computer
science~\cite{papa94,motwani95}
and form the basis for much computational physics,
such as the Monte Carlo simulations.
The recent flourish of interest in quantum computation and quantum
information~\cite{god,lee02a}
has lead to a number of studies of quantum random walks
both in continuous~\cite{farhi98,childs02a}
and in discrete time~\cite{aharonov93,meyer96,watrous01,aharonov01,ambainis01}.
Meyer has shown that a discrete time, discrete space, quantum random walk
requires an additional degree of freedom~\cite{meyer96},
or quantum ``coin,''
and can be modeled by a quantum lattice gas automaton~\cite{meyer01}.
Quantum random walks reveal a number of startling differences to their
classical counter parts.
In particular, the diffusion on a line
is quadratically faster.
Quantum random walks show promise as a means of implementing quantum algorithms.
Childs {\em et al}~\cite{childs02b} prove that a continuous time quantum random
walk
can find its way across some types of graphs exponentially faster than any
classical
random walk,
while a discrete time, coined quantum walk has been shown to equal Grover's
algorithm
in finding a specific item in an unsorted database with a quadratic speedup over
the best classical algorithm~\cite{shenvi03}.
A method of implementing a quantum random walk
in an ion trap computer has been proposed~\cite{travilgone01}.
A recent overview of quantum random walks is given by Kempe~\cite{kempe03}.
%Most work on discrete time, discrete space quantum random walks
%has concentrated on the unbiased walk driven by the Hadamard ``coin'':
%\begin{equation}
%\hat{H} = \frac{1}{\sqrt{2}} \begin{pmatrix}
%                               1 & 1 \\
%                               1 & -1
%                            \end{pmatrix}
%\end{equation}
%\egin{equation}
%\begin{split}
%|R\rangle &\rightarrow (|R\rangle + |L\rangle)/\sqrt{2} \\
%|L\rangle &\rightarrow (|R\rangle - |L\rangle)/\sqrt{2}
%\end{split}
%\end{equation}
%where $|R\rangle$ and $|L\rangle$ are the coin states.

Parrondo's games or Parrondo's paradox arises where a combination
of two losing games result in a winning game~\cite{harmer99a,mcclintock99}.
Such an effect can occur when one game has a form of feedback,
for example, through a dependence on the game state~\cite{harmer99b},
through the outcomes of previous games~\cite{parrondo00},
or through the states of neighbors~\cite{toral01},
that leads to a negative bias.
When this feedback is disrupted by mixing the play with a second losing game
that acts as a source of noise,
a net positive bias may result.
The recent attention attracted by classical versions of Parrondo's games
is motivated by their relation
to physical systems such as flashing ratchets
or Brownian motors~\cite{feynman63,harmer01,allison02},
or systems of interacting spins~\cite{moraal00}.
Applications in fields as diverse as
population genetics~\cite{mcclintock99},
biogenesis~\cite{davies01},
economics and biochemistry~\cite{klarreich01b} have been suggested.
Quantum equivalents to Parrondo's games
with a pay-off dependence~\cite{meyer01} or a history
dependence~\cite{ng01,flitney02}
have been demonstrated.
A link between quantum Parrondo's games and quantum algorithms
has been discussed~\cite{lee02b,lee02c}.
Recent reviews of classical and quantum Parrondo's games can be found in
Refs.~\cite{harmer02} and~\cite{flitney03}, respectively.
In this paper we develop a model of a quantum random walk
with history dependence
and detail its main features.
We show that this can lead to a new quantum version of Parrondo's paradox.

The paper is divided as follows.
Section~\ref{sec-parrondo} gives a brief summary of the classical Parrondo's
games and their quantum analogs,
Sec.~\ref{sec-formal} sets out the mathematical formalism of our scheme,
Sec.~\ref{sec-results} gives some results for the random walk of a single
particle on a line with this scheme,
while Sec.~\ref{sec-newparrondo} demonstrates a new quantum Parrondo effect.

\section{Parrondo's games}
\label{sec-parrondo}

The original Parrondo's games were cast in the form of a pair of gambling games,
game A the toss of a simple biased coin
with winning probability $p=\frac{1}{2} - \epsilon$,
and game B consisting of two biased coins,
the selection of which depends upon the state of the game.
Coin ${\rm  B}_1$, with winning probability $p_1$,
is selected when the capital is a multiple of three,
while coin ${\rm B}_2$, with winning probability $p_2$,
is chosen otherwise.
Each coin toss results in the gain or loss of one unit of capital.
With, for example,
\begin{equation}
\label{eqn-prob1}
p_1 = 1/10 - \epsilon,
\makebox[0.5cm]{} p_2 = 3/4 - \epsilon,
\makebox[0.5cm]{} \epsilon > 0,
\end{equation}
game B is losing since the ``bad'' coin ${\rm B}_1$ is played more often
than the one-third of the time that one would naively expect.
By interspersing plays of games A and B,
the probability of selecting ${\rm B}_1$ approaches $\frac{1}{3}$,
and that game produces a net positive result
that can more than offset the small loss from game A,
when $\epsilon$ is small.
The combination of the two losing games to form a winning one is the essence
of the apparent paradox first described by Parrondo.

\begin{figure}
\epsfig{file=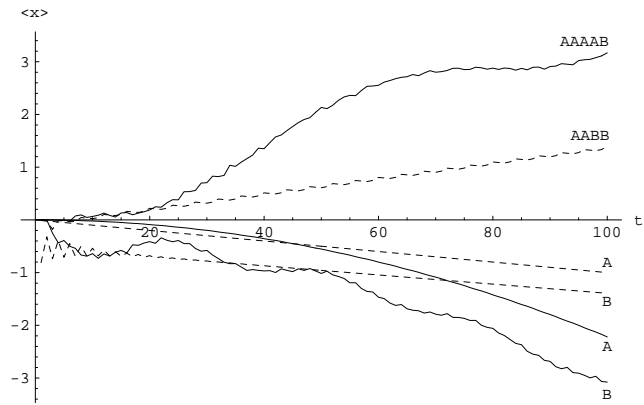, width=8.5cm}
\caption{Mean position $\langle x \rangle$ as a function of time
(in number of coin tosses)
for (dashed lines) the classical games A, B and the repeated sequence AABB
with $\epsilon = 0.005$ in Eq.~(\ref{eqn-prob1}),
and (solid lines) the quantum games A, B and the repeated sequence AAAAB
with $\alpha = \pi/2500$ and $\beta = \pi/3$ in Eq.~(\ref{eqn-potentials}).
In the classical case, $x$ is the player's capital
with \$1 awarded for each winning coin toss
and -\$1 for each losing toss.
Here, $x$ is the particle position
and we assume full coherence is maintained in the quantum case.
The difference in payoffs between the classical and quantum examples
is due to the particular parameters chosen.
However, interference in the quantum case produces a greater turn around in $x$
than is obtainable in the classical situation.}
\label{fig-oldparrondo}
\end{figure}

Meyer and Blumer~\cite{meyer01} were the first to present a quantum version
of this effect.
In their model, the quantum analog of the capital is
the discretization of the position of a particle undergoing Brownian motion
in one dimension.
Each play of the game changes the particle position by $\pm 1$ unit
in the $x$ direction.
The biases of game A and B are achieved by the application of potentials
\begin{equation}
\label{eqn-potentials}
\begin{split}
V_A(x) &= \alpha x, \quad \alpha > 0, \\
V_B(x) &= V_A(x) \:+\: \beta (1 - \frac{1}{2}(x \!\! \mod 3)), \quad \beta > 0,
\end{split}
\end{equation}
respectively.
By adjusting the parameters of the potentials,
the quantum games A and B can be made to yield similar negative
biases to their classical counter parts.
When switching between the potentials is introduced,
the bias can be reversed for certain mixtures of A and B.
For the classical and quantum versions,
comparisons of the expectations for the individual games
and an example of a winning combination
are given in Fig.~\ref{fig-oldparrondo}.
For details of the classical case
see Harmer and Abbott~\cite{harmer99b}
and for the quantum case Meyer and Blumer~\cite{meyer01}.

\begin{figure}
\begin{picture}(300,150)(25,25)
\put(130,140){\small results of previous two games}
\put(50,130){
\begin{picture}(270,20)(0,-5)
        \put(0,0){\line(1,0){270}}
        \multiput(0,0)(90,0){4}{\line(0,-1){10}}
\end{picture}}
\put(0,0){
\begin{picture}(90,120)(0,0)
        \put(29,110){\small{lost, lost}}
        \put(50,105){\line(0,-1){10}}
        \put(45,85){${\rm B}_1$}
        \put(35,50){\line(1,2){15}}
        \put(50,80){\line(1,-2){15}}
        \put(8,60){\small{$1-p_1$}}
        \put(68,60){\small{$p_1$}}
        \put(20,35){\small{lose}}
        \put(65,35){\small{win}}
\end{picture}}
\put(90,0){
\begin{picture}(90,120)(0,0)
        \put(29,110){\small{lost, won}}
        \put(50,105){\line(0,-1){10}}
        \put(45,85){${\rm B}_2$}
        \put(35,50){\line(1,2){15}}
        \put(50,80){\line(1,-2){15}}
        \put(10,60){\small{$1-p_2$}}
        \put(68,60){\small{$p_2$}}
        \put(20,35){\small{lose}}
        \put(65,35){\small{win}}
\end{picture}}
\put(180,0){
\begin{picture}(90,120)(0,0)
        \put(30,110){\small{won, lost}}
        \put(50,105){\line(0,-1){10}}
        \put(45,85){${\rm B}_3$}
        \put(35,50){\line(1,2){15}}
        \put(50,80){\line(1,-2){15}}
        \put(10,60){\small{$1-p_3$}}
        \put(68,60){\small{$p_3$}}
        \put(20,35){\small{lose}}
        \put(65,35){\small{win}}
\end{picture}}
\put(270,0){
\begin{picture}(90,120)(0,0)
        \put(30,110){\small{won, won}}
        \put(50,105){\line(0,-1){10}}
        \put(45,85){${\rm B}_4$}
        \put(35,50){\line(1,2){15}}
        \put(50,80){\line(1,-2){15}}
        \put(8,60){\small{$1-p_4$}}
        \put(68,60){\small{$p_4$}}
        \put(20,35){\small{lose}}
        \put(65,35){\small{win}}
\end{picture}}
\end{picture}
\caption{In the classical history-dependent Parrondo's game B,
the selection of coins ${\rm B}_1$ to ${\rm B}_4$
depends upon the results of the last two plays, as shown.
The probabilities of winning
(increasing the player's capital by one)
are $p_1$ to $p_4$
and of losing
(decreasing the player's capital by one)
are $1-p_1$ to $1-p_4$.
The overall payoff for a series of games is the player's final capital.}
\label{fig-history}
\end{figure}
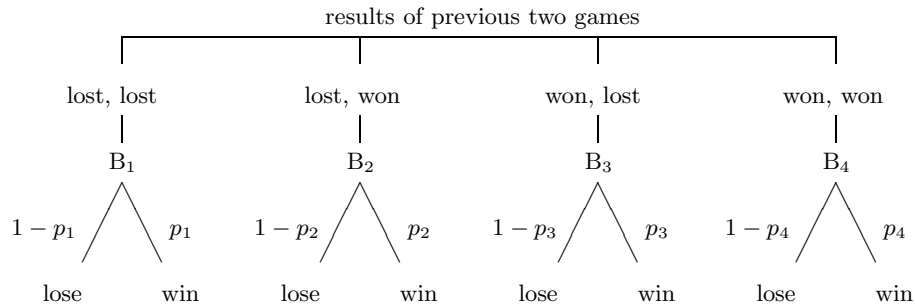

A history-dependent game
can be substituted for the above game B
to produce a variant of Parrondo's games.
Game B consists of four coins whose choice is determined by
the results of the previous two games,
as indicated in Fig.~\ref{fig-history}.
An analysis of this game for
\begin{equation}
p_1 = 7/10 - \epsilon,
\makebox[0.5cm]{} p_2 = p_3 = 1/4 - \epsilon,
\makebox[0.5cm]{} p_4 = 9/10 - \epsilon,
\end{equation}
indicates that the game is losing for $\epsilon > 0$~\cite{parrondo00}.
Mixing this with game A or a different history-dependent game
B~\cite{kay02} can yield an overall winning result.
A direct quantization of this scheme is given by
Flitney {\em et al}~\cite{flitney02}.
The quantum effects in this model depend upon the selection of a suitable
superposition as an initial state.
Interference can then arise since there may be more than one way
of obtaining a particular state.
Without interference, this scheme gives the same results as the classical
history-dependent Parrondo's game.
The method presented in the current article uses
an alternative approach,
a discrete quantum random walk or quantum lattice gas automaton.

\section{Scheme formalism}
\label{sec-formal}
\begin{figure}
\epsfig{file=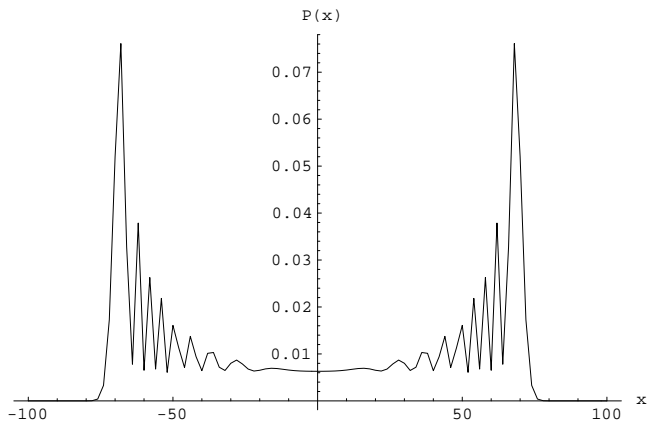,width=8.5cm}
\caption{The distribution of probability density
$P(x) = |\psi(x)|^2$ at toss $t=100$
for an unbiased, single coin quantum random walk
with $|\psi_0\rangle = \frac{1}{\sqrt{2}} (|0, L\rangle - |0, R\rangle)$.
Only even positions are plotted since $\psi(x)$ is zero for odd $x$ at $t=100$.
The total area under the graph is equal to one.}
\label{fig-unbiasM1}
\end{figure}

A direct translation of a classical discrete random walk into the quantum
domain is not possible.
If a quantum particle moving along a line
is updated at each step, in superposition, to the left and right,
the global process is necessarily non-unitary.
However, the addition of a second degree of freedom,
the chirality, taking values $L$ and $R$,
allows interesting quantum random walks to be constructed.
Consider a particle whose position is discretized in one-dimension.
Let ${\cal H}_P$ be the Hilbert space of particle positions,
spanned by the basis $\{|x\rangle : x \in \bf{Z}\}$.
In each time step the particle will move either to the left or right
depending on its chirality.
Let ${\cal H}_C$ be the Hilbert space of chirality, or ``coin'' states,
spanned by the orthonormal basis $\{|L\rangle, |R\rangle\}$.
A simple quantum random walk
in the Hilbert space ${\cal H}_P \otimes {\cal H}_C$
consists of a quantum mechanical ``coin toss,''
a unitary operation $\hat{U}$ on the coin state,
followed by the updating of the position to the left or right:
\begin{equation}
\hat{E} = (\hat{S} \otimes \hat{P}_R \;+\; \hat{S}^{-1} \otimes \hat{P}_L)
        (\hat{I}_P \otimes \hat{U}),
\end{equation}
where $\hat{S}$ is the shift operator in position space,
$\hat{S} |x\rangle = |x+1\rangle$,
$\hat{I}_P$ is the identity operator in position space,
and $\hat{P}_R$ and $\hat{P}_L$ are projection operators on the coin space
with $\hat{P}_R + \hat{P}_L = \hat{I}_C$, the coin identity operator.
For example, a walk controlled by an unbiased quantum coin
is carried out by the transformations
\begin{equation}
\label{eqn-step1}
\begin{split}
|x, L \rangle &\rightarrow\, \frac{1}{\sqrt{2}}
        \left( |x-1, L \rangle \,+\, i |x+1, R \rangle \right), \\
|x, R \rangle &\rightarrow\, \frac{1}{\sqrt{2}}
        \left(i |x-1, L \rangle \,+\, |x+1, R \rangle \right).
\end{split}
\end{equation}
Figure~\ref{fig-unbiasM1} shows the distribution of probability density
after 100 steps of Eq.~(\ref{eqn-step1}) with the initial state
$|\psi_0\rangle = (|0, L \rangle - |0, R \rangle)/\sqrt{2}~$\cite{note-initial}.
This initial state is chosen so that a symmetrical distribution results.
In fact the states $|0, R\rangle$ and $|0, L\rangle$ evolve independently.
We can see this since any flip $|R\rangle \leftrightarrow |L\rangle$
involves multiplication by a factor of $i$.
Thus, any $|x, L\rangle$ state that started from $|0, R\rangle$
will be multiplied by an odd power of $i$
and is orthogonal to any $|x, L\rangle$ state
that originated from $|0, L\rangle$
(and similarly for the $|x, R\rangle$ states).

To construct a quantum random walk with history dependence requires
an extension of the Hilbert space
by additional coin states.
Where we have a dependence on the last $M-1$ results,
the total system Hilbert space is a direct product between
the particle position in one dimension
and $M$ coin states:
\begin{equation}
{\cal H} = {\cal H}_P \otimes ({\cal H}_C  \: {}^{\otimes M}).
\end{equation}
The $M$ coins represent the results of tosses at times
$t-1, \, t-2, \ldots, t-M$.
A single step in the walk consists of tossing the $M$th coin,
updating the position
depending on the result of the toss,
and then re-ordering the coins
so that the newly tossed coin is in the first (most recent) position.
In general,
the unitary coin operator $\hat{U}$
can be specified,
up to an overall phase that is not observable,
by three parameters,
two of which are phases.
In the single coin case the effect of the phases can be completely
mimicked by changes to $|\psi_0\rangle$~\cite{tregenna03}.
This does not carry over to our multi-coin history-dependent scheme.
However, for the sake of simplicity we shall omit the phases and simply write
\begin{equation}
\label{eqn-flip}
\hat{U}(\rho) = \begin{pmatrix} \sqrt{\rho} & i \sqrt{1-\rho} \\
                                i \sqrt{1-\rho} & \sqrt{\rho}
                \end{pmatrix},
\end{equation}
where $1-\rho$ is the classical probability that the coin changes state,
with $\rho = \frac{1}{2}$ being an unbiased coin.
To allow for history dependence,
$\rho$ will depend upon
the results of the last $M-1$ coin tosses,
so that a single toss is effected by the operator
\begin{equation}
\label{eqn-Ehat}
%\begin{split}
\hat{E} = \left( \hat{S} \otimes \hat{I}_C {}^{\otimes (M-1)} \otimes \hat{P}_R
          \:+\: \hat{S}^{-1} \otimes \hat{I}_C {}^{\otimes (M-1)}
                \otimes \hat{P}_L \right)
          \left( \hat{I}_P \otimes \!\!\! \sum_{j_1,\ldots,j_{M-1} \in \{L,R\}}
                \!\!\! \hat{P}_{j_1 \ldots j_{M-1}}^{*} \otimes \hat{U}
                        (\rho_{j_1 \ldots j_{M-1}}) \right),
%\end{split}
\end{equation}
where $\hat{P}_j, \; j \in \{L,R\}$ is the projection operator
of the $M$th coin onto the state $|j\rangle$
and $\hat{P}_{j_{1} \ldots j_{M-1}}^{*}, \; j_k \in \{L,R\}$
is the projection operator of the
first $M-1$ coins onto the state $|j_{1} \ldots j_{M-1} \rangle$.
The second parenthesised term in (\ref{eqn-Ehat})
flips the $M$th coin with a parameter $\rho$
that depends upon the state of the first $M-1$ coins,
while the first term updates the particle position depending
on the result of the flip.
Re-ordering of the coins is then achieved by
\begin{equation}
\hat{O} = \hat{I}_P \otimes \!\!\! \sum_{j_1,\ldots,j_{M} \in \{L,R\}}
        \!\!\! |j_{M} j_{1} \ldots j_{M-1}\rangle \langle j_1 \ldots j_{M-1}
j_{M}|.
\end{equation}
This scheme is distinguished from Brun {\em et al}'s work
on quantum walks with multiple coins~\cite{brun02}
where the walk is carried out by cycling through
a given sequence of $M$ coins, $\hat{U}(\rho_1), \ldots, \hat{U}(\rho_M)$.
In Brun's scheme, a coin toss is performed by
\begin{equation}
%\begin{split}
\hat{E} = (\hat{S} \otimes \hat{I}_C {}^{\otimes (M-1)} \otimes \hat{P}_R
                \;+\; \hat{S}^{-1} \otimes \hat{I}_C {}^{\otimes (M-1)}
                \otimes \hat{P}_L)
          \left( \hat{I}_P \otimes \hat{I}_C {}^{\otimes (M-1)} \otimes \hat{U}
                (\rho_{k}) \right),
%\end{split}
\end{equation}
where $k = (t \! \mod M)$,
and the step is completed by the $\hat{O}$ operator as before.
The scheme has memory
but not the dependence on history
of the current method.
The two schemes are only equivalent when all
the $\rho_k$ and $\rho_{j_1 \ldots j_{M-1}}$ are equal,
for example, when all the coins are unbiased.
This amounts to asserting that the scheme of Brun {\em et al}
does not display Parrondian behavior.

\section{Results}
\label{sec-results}
\begin{figure}
\epsfig{file=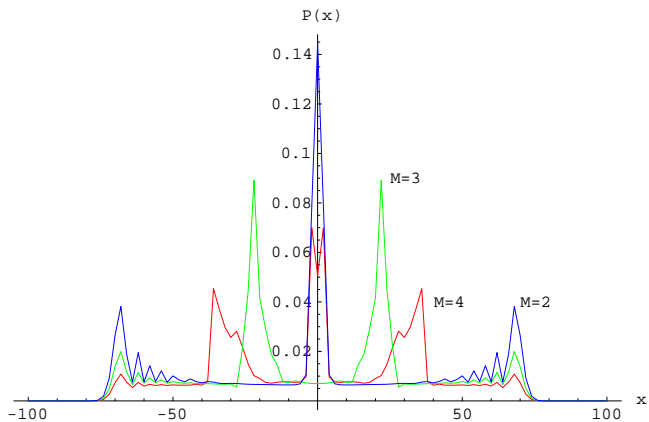,width=8.5cm}
\caption{(Color online)
The probability density distributions
$P(x) = |\psi(x)|^2$ at toss $t=100$,
for the 2- (blue), 3- (green) and 4- (red) coin unbiased, symmetrical,
quantum random walks.
Only even positions are plotted since $\psi(x)$ is zero for odd $x$ at $t=100$.
The area under each curve is equal to one.}
\label{fig-unbiasM234}
\end{figure}

The probability density distributions for
unbiased 2, 3, and 4 coin history-dependent quantum random walks,
with initial states that are an equal superposition of the possible coin states
antisymmetric as $L \leftrightarrow R$~\cite{note-antisym}
are shown in Fig.~\ref{fig-unbiasM234}.
These distributions are essentially symmetric versions of the graphs
of Brun {\em et al}~\cite{brun02} that result from an initial state
$|\psi_0\rangle = |R\rangle^{\otimes M}$.

For arbitrary $M$ we have,
as for the $M=1$ case,
two parts of the initial state that evolve without interacting.
Thus, for $M=2$ for example, states arising
from $|0, LL\rangle$ and $|0, RR\rangle$ will interfere,
as will states arising from $|0, LR\rangle$ and $|0, RL\rangle$,
but the two groups evolve into states that are orthogonal, for any given $x$.
For the $M$ coin quantum random walk there are $M+1$ peaks
with even values of $M$ having a central peak,
the others necessarily being symmetrically placed around $x=0$
by our choice of initial state.
The outer most pair of peaks are in the same position as the peaks for $M=1$
(Fig.~\ref{fig-unbiasM1})
at $x(t) \approx 0.68 t$.
All the peaks are interference phenomenon,
the central one being the easiest to understand.
It arises since there are states centred on $x=0$ that cycle back to themselves
(i.e., that are stationary states over a certain time period).
With $M=2$, the simplest cycle over $t=2$ is proportional to
\begin{equation}
\label{eqn-stationary}
\begin{split}
|0, LR\rangle \:-\: |0, RL\rangle &\rightarrow \frac{1}{\sqrt{2}}
        (|+ \!\! 1, RL\rangle \:+\: i |- \!\! 1, LL\rangle
                \:-\: |- \!\!1, LR\rangle \:-\: i |+ \!\! 1, RR\rangle) \\
 &\rightarrow |0, LR\rangle \:-\: |0, RL\rangle.
\end{split}
\end{equation}
At the second step,
complete destructive interference occurs for the states with $x = \pm 2$,
so that there is no probability flux leaving the central three $x$ values.
In practice, the central region asymptotically approaches a more
complex stationary cycle than (\ref{eqn-stationary}),
such as the $t=2$ cycle
\begin{equation}
\begin{split}
|\psi_{\rm center}\rangle &\propto (a i - b)(|-2, LL\rangle \:+\: |+2, RR\rangle)
        \:+\: (1 - a - i + b i) (|-2, LR\rangle \:+\: |+2,RL\rangle) \\
 & \makebox[5mm]{} +\: (i - 1)(|-2, RL\rangle \:+\: |+2, LR\rangle)
        \:+\: (b - a i)(|0, LL\rangle \:+\: |0, RR\rangle)
        \:+\: (a + b i)(|0, LR\rangle \:+\: |0, RL\rangle),
\end{split}
\end{equation}
where $a$ and $b$ are real.

\begin{figure}
\epsfig{file=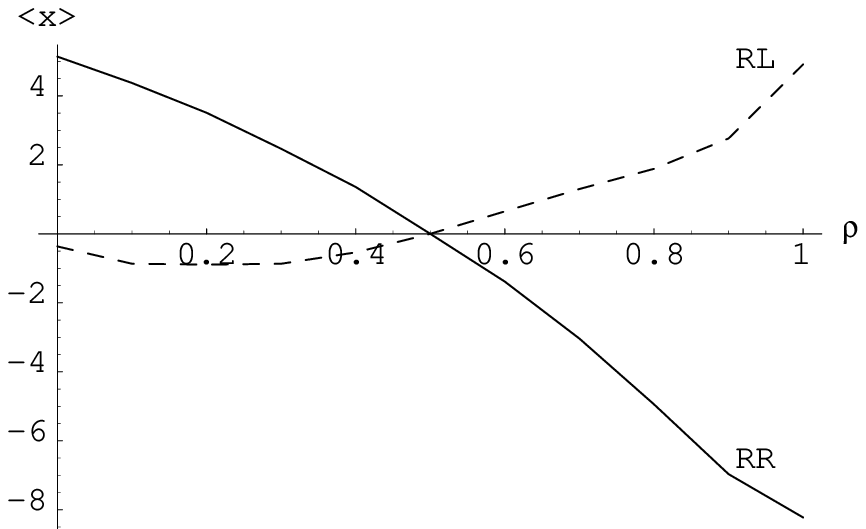,width=8.5cm}
\epsfig{file=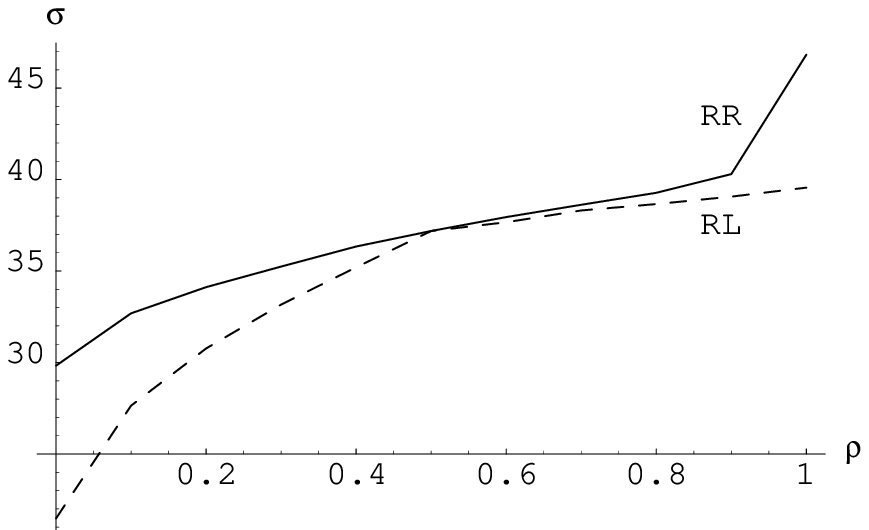,width=8.5cm}
\caption{For the $M=3$ quantum history-dependent walk,
$\langle x \rangle$ and $\sigma_x$ at time step $t=100$
as a function of $\rho_{RR}$ (solid line) or $\rho_{RL}$ (dashed line)
while the other $\rho_{ij}$ are kept constant at 1/2.
Varying $\rho_{LL}$ has the opposite effect on $\langle x \rangle$
and the same on $\sigma_x$ as varying $\rho_{RR}$.
Similarly for $\rho_{LR}$ compared to $\rho_{RL}$.}
\label{fig-bias}
\end{figure}

Adjusting the values of the various $\rho$ can introduce a bias into the walk.
To create a quantum walk analogous to the history-dependent game B of
Sec.~\ref{sec-parrondo}, requires $M=3$,
giving four parameters,
$\rho_{RR}, \, \rho_{RL}, \, \rho_{LR}$ and $\rho_{LL}$.
Figure~\ref{fig-bias} shows the effect of individual variations
in these parameters on the expectation value and standard deviation of the position
after 100 time steps.


\section{Quantum Parrondo effect}
\label{sec-newparrondo}
It is useful to consider the classical limit to our quantum scheme.
That is, the random walk that would result
if the scattering amplitudes were replaced by classical probabilities.
As an example consider the $M=2$ case,
with winning probabilities $1-\rho_L$ and $1-\rho_R$.
The analysis below follows that of Harmer and Abbott~\cite{harmer02}.
Markov chain methods cannot be used directly
because of the history dependence of the scheme.
If, however, we form the vector
\begin{equation}
y(t) = [ x(t-1)-x(t-2), \; x(t)-x(t-1) ],
\end{equation}
where $x(t)$ is the position at time $t$,
then $y(t)$ forms a discrete time Markov chain between the states
$[-1,-1],\, [-1,+1],\, [+1,-1]$ and $[+1,+1]$ with a transition matrix
\begin{equation}
T = \begin{pmatrix}
        \rho_L &  1-\rho_L & 0 & 0 \\
        0 & 0 & \rho_R & 1-\rho_R \\
        1-\rho_L & \rho_L & 0 & 0 \\
        0 & 0 & 1-\rho_R & \rho_R
    \end{pmatrix}
\end{equation}
Define $\pi_{ij}(t)$ to be the probability of $y(t)=[i,j], \; i,j \in \{-1,+1\}$.
A state is now transformed by $T {\bm \pi}$ at each time step.
Having represented the history-dependent game as a discrete time Markov chain
the standard Markov techniques can be applied.
The equilibrium distribution is found by solving $T {\bm \pi_s} = {\bm \pi_s}$.
This yields ${\bm \pi_s} = [1,1,1,1]/4$,
giving a process with no net bias to the left or right
irrespective of the values of $\rho_L$ and $\rho_R$.
The same analysis holds for $M>2$.
However, interference in the quantum case presents an entirely different picture.

\begin{figure}
\epsfig{file=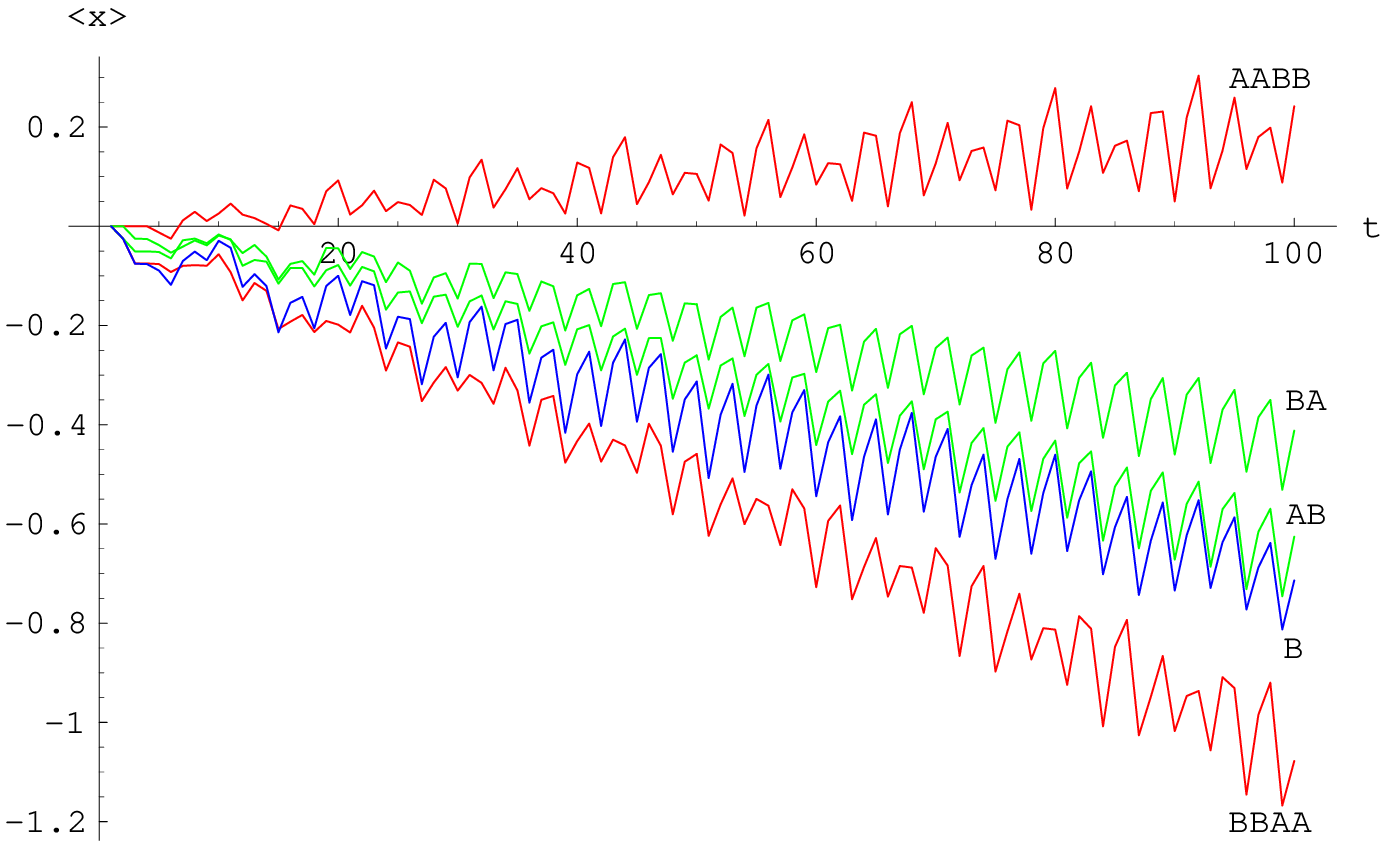,width=8.5cm}
\begin{picture}(270,40)(0,0)
\end{picture}
\epsfig{file=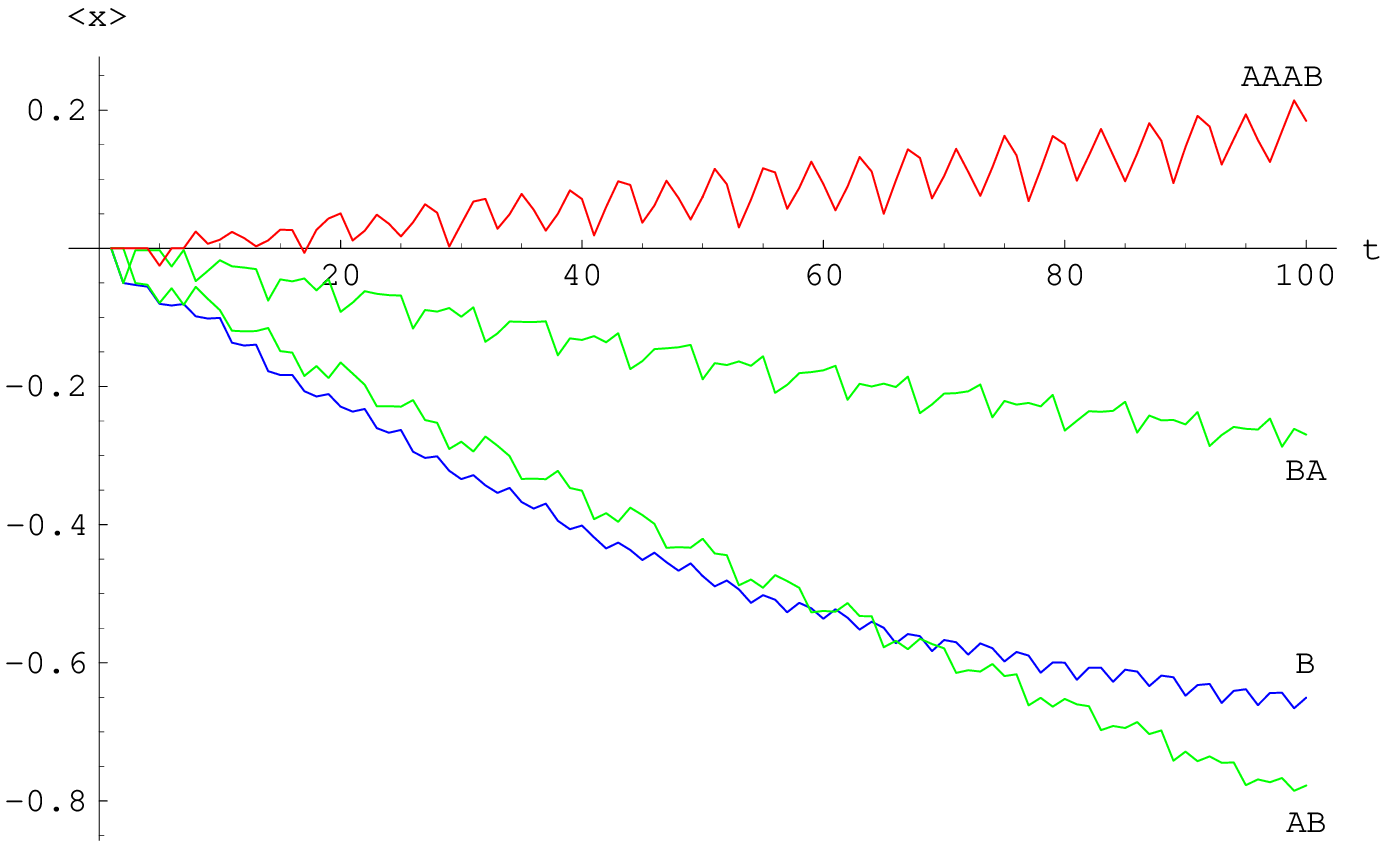,width=8.5cm}
\caption{(Color online)
An example of a Parrondo effect for the $M=3$ history-dependent quantum
random walk where game B has (top) $\rho_{RR} = 0.55$
or (bottom) $\rho_{LR} = 0.6$,
with the other $\rho_{ij} = 0.5, \; i,j \in \{L,R\}$,
while game A has all $\rho_{ij} = 0.5$ (unbiased).
The letters next to each curve represent the sequence of games
played repeatedly.
For example, AB means apply $\hat{A}$ and then $\hat{B}$ to the state,
repeating this sequence 50 times to get to $t=100$.}
\label{fig-newparrondo}
\end{figure}

The comparison with the classical history-dependent Parrondo game
requires $M=3$.
For game A, select the unbiased game,
$\rho_{LL} = \rho_{LR} = \rho_{RL} = \rho_{RR} = 1/2$.
For game B, choose, for example, $\rho_{RR} = 0.55$ or $\rho_{LR} = 0.6$
to produce a suitable bias
(see Fig.~\ref{fig-bias}).
The operators associated with A and B are applied repeatedly,
in some fixed sequence,
to the state $|\psi\rangle$.
For example, the results of the game sequence AABB after $t$ time steps is
\begin{equation}
|\psi(t)\rangle = (\hat{B} \hat{B} \hat{A} \hat{A})^{t/4} |\psi(0)\rangle.
\end{equation}
Figure~\ref{fig-newparrondo} displays $\langle x \rangle$ for various sequences.
Of sequences up to length four,
with game B biased by $\rho_{RR} > 0.5$
only AABB and AAB give a positive expectation,
while when game B is biased by $\rho_{LR} > 0.5$
only AAAB is positive.
These results hold for $\rho$ up to approximately 0.6,
above which there are no positive sequences
of length less than or equal to four.

The sequences AABB and BBAA can be considered the same but with different
initial states.
That is, if instead of $|\psi_0\rangle$,
we start with $|\psi_0'\rangle = \hat{A} \hat{A} |\psi_0\rangle$,
BBAA gives the same results (displaced by two time steps) as AABB
does with the original starting state.
In the classical case,
altering the order of the sequence
results in the same trend but with a small offset, as one might expect.
However, as Fig.~\ref{fig-newparrondo} indicates,
the change of order in the quantum case can produce radically different results.
This feature also appears in Meyer and Blumer's quantum Parrondo model.

\section{Conclusion}
A scheme for a discrete quantum random walk with history dependence
has been presented.
Our system involves the use of multiple quantum coins.
By suitable selection of the amplitudes for coin flips dependent
on certain histories,
the walk can be biased to give positive or negative $\langle x \rangle$.
In common with many other properties of quantum random walks,
the bias results from interference,
since the classical equivalent of our random walks are unbiased.
With a starting state averaged over possible histories,
the average spread of probability density in our multi-coin scheme
is slower than in the single coin case,
with the appearance of multiple peaks in the distribution.
For even numbers of coins there is a substantial probability of $x \approx 0$.
However, the positions of the outer most peaks
are the same as those of a single coin quantum random walk.
As the memory effect increases,
the dispersion of the quantum walk decreases.
One may speculate that this feature may be relevant
to an understanding of decoherence,
here considered as loss of coherence within the central portion of the graph
around $x \approx 0$.
In particular, the dispersion in the wavefunction
decreases as we move from a first-order Markov system
to a non-first-order Markov system,
that is, one with memory.
This is consistent with the idea that the Markovian approximations
tend to over-estimate the decoherence of the system.

Our scheme is the quantum analog of the history-dependent game
in a form of Parrondo's paradox.
The quantum history-dependent walk also exhibits a Parrondo effect,
where the disruption of the history dependence in a biased walk
by mixing with a second, unbiased walk can reverse the bias.
In distinction to the classical case,
the effect seen here is very sensitive to the exact sequence of operations,
a quality it shares with other forms of quantum Parrondo's games.
This sensitivity is consistent with the idea that the effect relies
on full coherence over space and in time.

We have only considered a quantum walk on a line.
The effect of memory driven quantum walks on networks with different topologies
and whether the memory structure can be chosen to optimize the path
in such networks, are open questions.

\acknowledgments
AF would like to thank Luis Quiroga for useful discussions,
and Roland Kay and Alexandra Olaya-Castro for their support and friendship
during his visit to Oxford.
This work was supported by GTECH Corporation Australia
with the assistance of the SA Lotteries Commission (Australia).
Travel assistance for one of us (AF) was provided by
the D. R. Stranks travel fellowship
and by The University of Adelaide postgraduate travel award.


\begin{thebibliography}{99}
\bibitem{note-initial}
This scheme is equivalent to the Hadamard quantum random walk
with initial state $\frac{1}{\sqrt{2}} (|0, L\rangle + i |0, R\rangle)$.
\bibitem{note-antisym}
For example, with $M=2$, the initial state is
$|\psi_0\rangle = (|0, LL\rangle - |0, LR\rangle - |0, RL\rangle + |0, LL\rangle)/2$.
For the purposes of this paper
we could equally well have chosen an initial state
that was symmetrical as $L \leftrightarrow R$,
however, the antisymmetric starting state is the one that gives the correct
behavior in the presence of a potential.
The state $|\psi_0\rangle$ is the quantum equivalent of the average over past histories
that is taken in the classical history-dependent Parrondo game.
\end{thebibliography}
\end{document}